\documentclass[]{spie}  

\usepackage{amsmath,amsfonts,amssymb}
\usepackage{graphicx}
\usepackage[colorlinks=true, allcolors=blue]{hyperref}
\usepackage{romannum}

\title{Adjoint method and inverse design for diffractive beam splitters}

\author[a,b]{Dong Cheon Kim}
\author[a]{Andreas Hermerschmidt}
\author[a]{Pavel Dyachenko}
\author[b]{Toralf Scharf}
\affil[a]{Holoeye Photonics AG, 12489 Berlin, Germany}
\affil[b]{Nanophotonics and Metrology Laboratory, \'Ecole Polytechnique F\'ed\'erale de Lausanne (EPFL),1015 Lausanne, Switzerland}

\authorinfo{Further author information: (Send correspondence to D.C.Kim)\\D.C.Kim: E-mail: dong.kim@epfl.ch, Telephone: +41 21 693 78 64}

\begin{document} 
\noindent \copyright (2020) COPYRIGHT Society of Photo-Optical Instrumentation Engineers (SPIE).

\noindent \textbf{Please Note:} According to SPIE Article-Sharing Policies “Authors may post draft manuscripts on preprint servers such as arXiv. If the full citation and Digital Object Identifier (DOI) are known, authors are encouraged to add this information to the preprint record.”\\
(\url{https://www.spiedigitallibrary.org/article-sharing-policies})

\noindent This document represents a draft from \textbf{ Dong Cheon Kim, Andreas Hermerschmidt, Pavel Dyachenko, and Toralf Scharf "Adjoint method and inverse design for diffractive beam splitters"}, submitted in Components and Packaging for Laser Systems \Romannum{6}, SPIE LASE 2020.

\maketitle

\begin{abstract}
Diffractive optical elements with a large diffraction angle require feature sizes down to sub-wavelength dimensions, which require a rigorous electromagnetic computational model for calculation.
However, the computational optimization of these diffractive elements is often limited by the large number of design parameters, making parametric optimization practically impossible due to large computation times. 
The adjoint method allows calculating the gradient of the target function with respect to all design variables with only two electromagnetic simulations, thus enabling gradient optimization.
Here, we present the adjoint method for modeling wide-angle diffractive optical elements like $7$x$7$ beam splitters with a maximum $53^\circ$ diffraction angle and a non-square $5$x$7$ array generating beam splitter.
After optimization we obtained beam splitter designs with a uniformity error of $16.35\%$ ($7$x$7$) and $6.98\%$ ($5$x$7$), respectively.
After reviewing the experimental results obtained from fabricated elements based on our designs, we found that the adjoint optimization method is an excellent and fast method to design wide-angle diffractive fan-out beam-splitters.
\end{abstract}

\section{INTRODUCTION}
\label{sec:intro}  
Diffractive optical elements (DOEs) are used for a variety of applications due to their high design flexibility, compact size, and mass productivity.
DOEs become key components for laser-based systems by controlling the shape of the beam.
Good examples of this type of devices are diffractive optical beam-splitting elements, often also referred to as fan-out gratings.
They create an array of the optical beam by deflecting an incident light into different diffraction orders at precise angles. 
Such elements therefore have various applications, ranging from multifocal microscopy\cite{Jureller2006, Chen2019}, optical interconnects\cite{Brosseau2000}, camera calibration\cite{Bauer2008} to structured light projectors\cite{Vandenhouten2017,Barlev2018}.
For many applications, the goal is to propagate to all the light in the desired orders and avoid losses.

To quantify the amount of light that propagates in the desired diffraction orders, we define the diffraction efficiency $\eta$ as the ratio of the sum of the powers of all desired orders to the total power of the incident beam.
\begin{align}\label{eq:de}
\eta = \frac{P_{out}}{P_{in}}
\end{align}
In addition, the diffractive beam splitters are usually designed to create an array of equal-intensity.
The uniformity of the spot array output is also the primary concern and is usually evaluated by considering the uniformity error (UE) and root-mean-square error (RMSE) $\sigma$:
\begin{align}\label{eq:ue}
UE = \frac{\eta_{max}-\eta_{min}}{\eta_{max}+\eta_{min}}
\end{align}
\begin{align}\label{eq:rmse}
\sigma=\sqrt{\frac{1}{N}\sum_{m=-M}^{M}(\eta_{m}-\hat{\eta})^2}    
\end{align}
where $\eta_{max}$ and $\eta_{min}$ represent the maximal diffraction efficiency and the weakest one in orders. The target diffraction efficiency $\hat{\eta}$, e.g., an average diffraction efficiency, and the $\eta_{m}$ is the diffraction efficiency in each order and $N$ is the total number of diffraction orders from $-M$ to $M$.

The angular separation of the diffraction orders is determined by the grating period and the wavelength of the incident light, while the distribution of light between the various diffraction orders is determined by the features of the structure within a single grating period. 
Since the diffraction angle is proportional to the ratio of the wavelength of the incident light to the grating period, a grating is often shrunk to create a wide-angle spot pattern. 
However, when the size of structures in the gratings becomes comparable with the wavelength of the incident light, the iterative Fourier transform algorithm (IFTA) \cite{SKEREN2002} based on thin element approach, which is one of the most popular methods, cannot be applied to optimize the grating because rigorous methods have to be applied.
Thus, optical elements in the non-paraxial domain, i.e. the elements with wavelength-scale features, need simulations of their electromagnetic properties using rigorous diffraction theory, such as rigorous coupled-wave analysis (RCWA) \cite{Lalanne1996,Moharam1995,Liu2012}.

To design and optimize optical structures via numerical simulations, we often use the gradient of some figure of merit with respect to our design variables. 
The use of parametric optimization based on the rigorous analysis is often computationally heavy because the gratings with many parameters are leading to high-dimensional optimization problems.
With an adjoint variable method\cite{Lalau-Keraly2013,Piggott2015,Hughes2017,Wang2018}, one can calculate the gradient using only two simulations, no matter how many variables are present.
This capability allows for efficient, large-scale, gradient-based optimization of optical structures in the non-paraxial domain.

In this work, we focus on solving an inverse diffraction problem arising from the parameter reconstruction using the gradient of the figure of merit based on the adjoint method.
This paper is structured as follows: The adjoint method with RCWA is presented in Section \ref{sec:method}. Numerical results of the optimization and experimental results are presented in Section \ref{sec:result}.

\section{OPTIMIZATION METHOD}
\label{sec:method}
\begin{figure} [htbp]
   \begin{center}
   \begin{tabular}{c} 
   \includegraphics[width=0.95\textwidth]{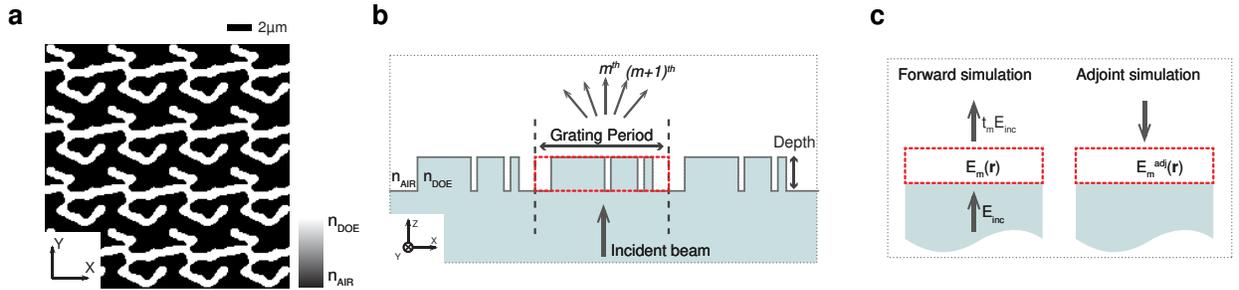}
   \end{tabular}
   \end{center}
   \caption[geometry]{ 
   \label{fig:geometry} 
Overview of diffractive beam splitter design using adjoint method. (a) Top view of a two-dimensional binary grating designed using adjoint-based optimization. Black represents air and white represents fused silica. (b) The surface profile of a two-dimensional grating structure. (c) Schematic of the forward and adjoint simulations used to optimize a large-angle diffractive beam splitter, which deflects normally incident light into the (0, 0) diffraction order.}
   \end{figure}

Figure \ref{fig:geometry}(a) and (b) show an example of a two-dimensional binary phase grating profiles characterized by depth, grating period, refractive index distribution.
Despite the advancement in lithography technologies, high cost and fabrication errors are inevitable for the fabrication of multilevel gratings.
The binary grating is considerably easier to fabricate than multilevel or continuous-relief elements. 
In this study, we therefore use two-dimensional binary gratings for applying the optimization method.

\subsection{Adjoint variable method}
An important aspect of the design process is the parametrization used to describe the shape of the optical elements.
For the optimization, we assume the gratings consisting of pixels of relative permittivities corresponding to air ($\epsilon_{air}$) and dielectric material ($\epsilon_{dielectric}$).
This permittivity distribution $\mathbf{\epsilon(r)}$ in our grating area are design variables of the optimization.
The position vector $\mathbf{r}$ is in our design domain which is the $xy$ space of a single grating period.
Over the multiple iterations, the pixels of relative permittivities are conversing toward $\epsilon_{air}$ or $\epsilon_{dielectric}$ to optimize our figure of merit. 
We define the figure of merit ($\mathit{FoM}$), often referred to as target function, as standard deviation
\begin{align}\label{eq:fom}
F = \sum_{m=-M}^{M}(\eta_{m}-\hat{\eta})^2
\end{align}
where figure of merit $F(\mathbf{\epsilon(r)})$ that represents the difference between the calculated diffraction efficiency $\eta_{m}(\mathbf{\epsilon(r)})$ in orders and the average diffraction efficiency $ \hat{\eta}$. 
The gradient of figure of merit with respect to $\mathbf{\epsilon(r)}$ is crucial in determining the search direction to optima, which is the set of derivatives of figure of merit with respect to each pixel, i.e.
\begin{align} \label{eq:gradient}
\nabla F=\left[\frac{\partial F}{{\partial \epsilon \left (r_1 \right )}},\cdots,\frac{\partial F}{{\partial  \epsilon(r_k)}},\cdots,\frac{\partial F}{{\partial \epsilon(r_K)}}\right].
\end{align}
For instance, if the total pixel number $K$ is $2500$ ($50$x$50$ pixels), it may easily become computationally heavy to calculate the gradient in rigorous analysis.
The adjoint method however allows computing the variation for a figure of merit by only two rigorous simulations, no matter how many pixels are present.
To obtain the gradient using the adjoint method, Maxwell's equation is solved the first time for a given illumination by an RCWA solver. 
And one has to solve the adjoint simulation which is a kind of influence function regarding the figure of merit using dipole representation and Lorentz reciprocity.(see Figure \ref{fig:geometry}c).

The diffraction efficiencies $\eta_m$ can be obtained by considering the transmitted power flow going to the diffraction order represented by plane wave $\mathbf{E_m},\mathbf{H_m}$ : 
\begin{align}\label{eq:powerflow}
\eta_m=\left | t^2 \right |=\left | \iint \left \{ \mathbf{E}(z_0) \times \mathbf{H_m^-}(z_0) - \mathbf{E_m^-}(z_0) \times  \mathbf{H}(z_0) \right \}\cdot \mathbf{n_z}  d\mathbf{x} \right |^2
\end{align}
where both fields are evaluated at the $z=z_0$ plane above the grating, and the overlap integral is performed for a single grating period. The $k$-vector of $\mathbf{E_m},\mathbf{H_m}$ is $(k_x,k_y,k_z)$ and $k$-vector of $\mathbf{E_m^-},\mathbf{H_m^-}$ is $(k_x,k_y,-k_z)$ and the normalization is
$\left | \left ( \mathbf{E_m} \times \mathbf{H_m^-} - \mathbf{E_m^-} \times  \mathbf{H_m} \right ) \cdot \mathbf{n_z} \right |  = 1$.
The variation of $\eta_m$ for a small perturbation in permittivity $\Delta \epsilon$ at a location $r$ in the grating layer is given by:
\begin{align}\label{eq:deltaDE}
\delta\eta_m = 2  \Re\left [ t^\ast 
 \iint \left \{ \delta\mathbf{E}(z_0) \times \mathbf{H_m^-}(z_0) - \mathbf{E_m^-}(z_0) \times  \delta\mathbf{H}(z_0) \right \}\cdot \mathbf{n_z}  d\mathbf{x} \right ].
\end{align}
Using dipole representation and the Green's tensor, the derivative of diffraction efficiency with respect to permittivity is:
\begin{align}
\frac{\partial \eta_m }{\partial \epsilon}=  2 \Re\left \{ t^\ast \mathbf{E_{adj}(r)}  \cdot \mathbf{E(r)} \right \} 
\label{eq:adjoint}
\end{align}
where adjoint field $\mathbf{E_{adj}(r)}$ can be obtained by an auxiliary RCWA simulation with illumination condition which is a plane wave generated by the polarization and magnetization densities from the dipole expression in Fig \ref{fig:geometry}c.
The Eq. \ref{eq:adjoint} shows that one requires the gradient from only two simulations, one direct($\mathbf{E(r)}$) and one adjoint($\mathbf{E_{adj}(r)}$) to evaluate the derivatives at all pixels in a grating.
These gradients are used with the quasi-Newton method, i.e., Limited-memory Broyden-Fletcher-Goldfarb-Shanno(L-BFGS) algorithm \cite{Zhu1997} to obtain fast and stable convergence.
Constraints are added to the optimization to ensure that the final optimized design converges to a binary grating comprising only two relative permittivities, which are air and dielectric material.

\subsection{Binarization and maintaining minimum feature size}
\label{subsec:blurfilter}
To create realistic gratings with sufficiently large minimum feature sizes and binarized permittivity distributions, we employ filtering and projection functions during the optimization.
In order to apply these functions, we choose to update a design density $\mathbf{\rho}$ which varies between $0$ and $1$ rather than updating the permittivity distribution directly within the design region.
 For generating a structure with larger feature sizes a low pass filter can be applied to $\mathbf{\rho}$ to create a filtered density, labeled $\mathbf{\tilde{\rho}}$.
This defines a low-pass spatial filter on $\mathbf{\rho}$ with the effect of smoothing out features with length scale below filter radius $R$. 
Now, for binarization of the structure, a projection function is used to recreate the final permittivity from the filtered density. We define $\bar{\rho}$ as the projected density, which is created from $\tilde{\rho}$ as controlling the threshold of the projection and the strength of the projection.
we can describe an analytical solution of the determination of $\frac{\partial \mathbf{\epsilon}}{\partial \mathbf{\bar{\rho}}}$, $\frac{\partial \mathbf{\bar{\rho}}}{\partial \mathbf{\tilde{\rho}}}$, $\frac{\partial \mathbf{\tilde{\rho}}}{\partial \mathbf{{\rho}}}$, these filters can be combined with the derivatives of the figure of merit calculated by the adjoint method. 
These algorithms are part of the design process.

\section{OPTIMIZATION RESULTS}
\label{sec:result}
To apply the optimization method, we prepared two kinds of two-dimensional fan-out gratings calculated by IFTA for initial designs.
One is a square multi-spot generator which creates $7$x$7$ array of spots with equal-intensity, the other generates $5$x$7$ non-square array of spots.
Here, we only consider beam splitter generating the uniform distribution for the sake of simplicity, we however can also design non-uniform beam splitter with any distribution meeting the application's requirements.

\begin{figure} [htbp]
   \begin{center}
   \begin{tabular}{c} 
   \includegraphics[width=0.95\textwidth]{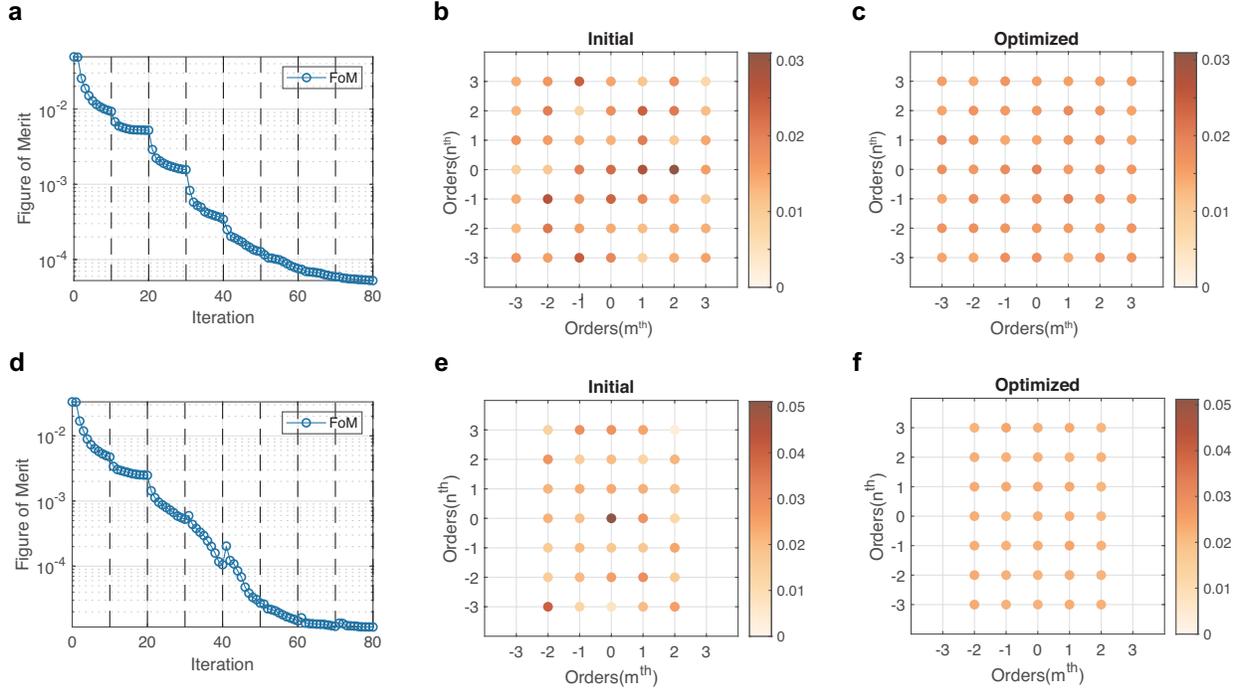}
   \end{tabular}
   \end{center}
   \caption[result1]{ 
   \label{fig:result1} 
Theoretical analysis of $7$x$7$ and $5$x$7$ diffractive beam splitter. Plot of figure of merit over the course of the adjoint-based optimization process for (a) the $7$x$7$ beam splitter and (d) the $5$x$7$ beam splitter. The calculated diffraction efficiency of $7$x$7$ beam splitter (b) before and (c) after optimization and the results of $5$x$7$ beam splitter (e) before and (f) after optimization.}
   \end{figure} 

\subsection{Reconstruction results}

\begin{table}[htbp]
\caption{Comparsion with the properties of the beam splitters before and after optimization.} 
\label{tab:results1}
\begin{center}       
\begin{tabular}{|l|p{1.8cm}|p{1.8cm}||p{1.8cm}|p{1.8cm}|} 
\hline
\rule[-1ex]{0pt}{3.5ex} {}&\multicolumn{2}{l||}{7x7 beam splitter}&\multicolumn{2}{l|}{5x7 beam splitter}   \\
\cline{2-5}
\rule[-1ex]{0pt}{3.5ex}   & Initial & Optimized & Initial & Optimized\\
\hline
\rule[-1ex]{0pt}{3.5ex}  Total efficiency($\%$) & 79.96  & 79.71 & 74.45  & 78.48   \\
\hline
\rule[-1ex]{0pt}{3.5ex}  UE($\%$) & 63.79 & 16.35 & 87.18 & 06.98    \\
\hline
\rule[-1ex]{0pt}{3.5ex}  UE in off-axis($\%$) & 63.79 & 15.63 & 83.95 & 06.98   \\
\hline
\rule[-1ex]{0pt}{3.5ex}  RMSE($\%$) & 0.533 & 0.126 & 0.852 & 0.075 \\
\hline
\end{tabular}
\end{center}
\end{table}   

The two-dimensional fan-out gratings with binary surface profiles have a $1.18 \mu m$ depth and $5$x$5 \mu m$ grating period.
The small pixels have a pixel size of only $100$x$100 nm$. 
The refractive index of a dielectric layer is $1.451$ for fused silica and $1.0$ for that of the air layer.
Starting from a profile designed by IFTA, a Limited-memory Broyden-Fletcher-Goldfarb-Shanno(L-BFGS) algorithm with the gradient calculated by the adjoint method was performed to obtain a low RMSE.
During the optimization, the diffraction efficiency of devices was evaluated always with the RCWA solver RETICOLO \cite{reticolo}.
Figure \ref{fig:result1}(a) shows the merit function as a function of the number of performed optimization iterations. 
The figure of merit converged well and the algorithm found the local optimum after 80 iterations.
The simulated diffraction efficiency of devices before and after optimization is shown in Fig.\ref{fig:result1}(b) and (c), respectively.
These results are calculated for normally incident TE polarized light, i.e. the electric field component along the y-axis.
In the diffraction pattern, the maximal diffraction angle is about $53^{\circ}$ at $(3,3)^{th}$ order spot from the center.
For an accurate comparison, we calculated the total diffraction efficiency, UE, UE in off-axis (i.e., excluding the zero-order), and RMSE of two different devices: Initial and optimized one.
These values are represented in Tab.\ref{tab:results1}.
The optimized device has no degradation in total efficiency compared to that of the initial one.
Through adjoint-based optimization there is considerable improvement in UE from $63.79\%$ to $16.35\%$ and RMSE from $0.533$ to $0.126$.

Furthermore, we applied this algorithm to optimize the beam-splitters creating non-square arrays.
One of the optimized results is shown in Fig. \ref{fig:result1}(f).
The convergence plot over the optimization of $5$x$7$ beam splitter is presented in Fig.\ref{fig:result1}(d).
A total of around 80 iterations are used to design the devices.
The sharp dips represent a strong geometric change that is applied every 10 iterations to eliminate small feature size and push to binarized distribution (see Sec.\ref{subsec:blurfilter}).
Over the course of multiple iterations, the dielectric continuum in the device converges to the dielectric constant of either silica or air.
The simulated diffraction pattern from initial dielectric distribution and the final one after optimization is presented in  Fig. \ref{fig:result1}(e) and (f), respectively.
We can observe the diffraction pattern distribution of optimized design is a nearly identical intensity to every target spot.
To evaluate the properties of $5$x$7$ beam splitter, we calculated the total diffraction efficiency, UE in off-axis, and RMSE~(see Table \ref{tab:results1}).
Through adjoint-based optimization, the total efficiency slightly increased from $74.45\%$ to $78.48\%$  and UE and RMSE consequently reached $6.98\%$ and $0.075$.
These results prove that the optimization algorithm is suitable for designing diffractive beam splitters.
Based on the optimized designs, we fabricated and characterized the mentioned beam splitters. The detailed experimental results are presented in the next section.

\subsection{Experimental results}
\begin{figure} [htbp]
   \begin{center}
   \begin{tabular}{c} 
   \includegraphics[width=0.95\textwidth]{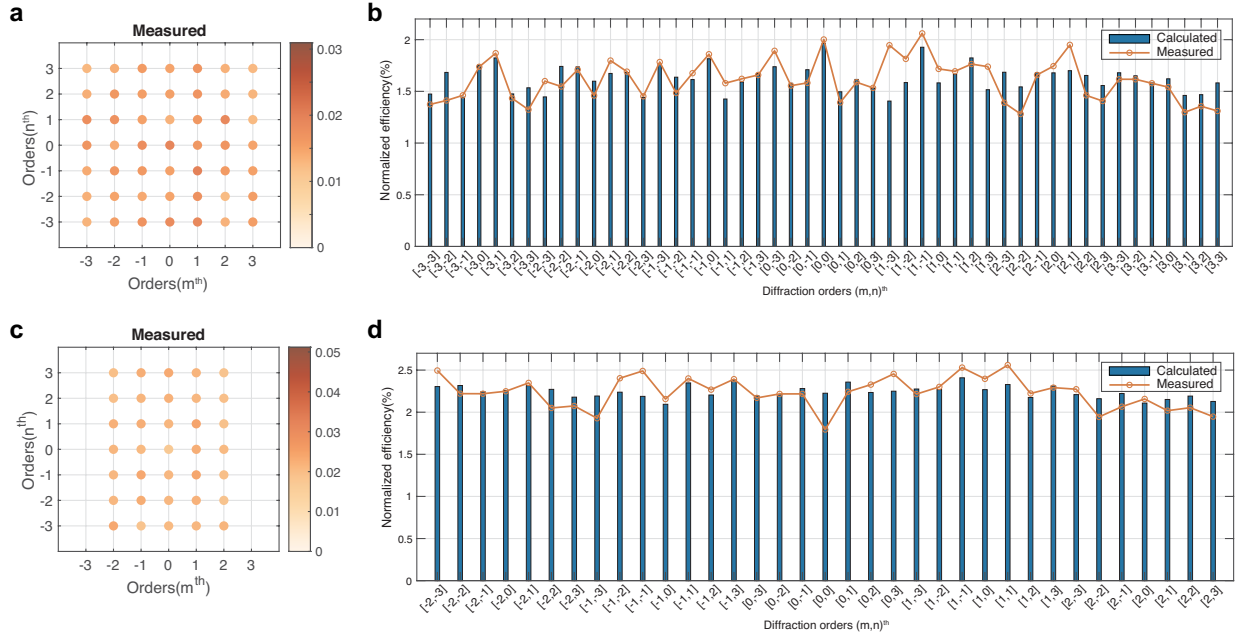}
   \end{tabular}
   \end{center}
   \caption[measured]{ 
   \label{fig:experimental} 
Experimental characterization of fan-out grating devices. The measured efficiency of (a) 7x7 beam splitter and (b) $5$x$7$ beam splitter. The experimental (orange line) and simulated efficiency (blue bar) of (c) 7x7 beam splitter and (d) $5$x$7$ beam splitter.}
   \end{figure}

The diffractive beam splitters were fabricated by lithography using electron-beam and dry etching to create a chromium etch mask, and then by reactive ion etching to obtain fused silica binary surface relief structures.
The devices are optically characterized using a $940 nm$ wavelength laser as our input source.
We detect the diffracted light beams using a mobile single-pixel detector with a high dynamic range.
Experimental diffraction efficiencies of $7$x$7$ beam splitter and $5$x$7$ beam splitter are shown in Fig. \ref{fig:experimental}(a) and (c), respectively.
The experimental data show that these devices operate with good uniformity which is close to the theoretical values.
The $7$x$7$ beam splitter sample has a total efficiency of  $76.67 \%$, UE of $22.00 \%$, UE in off-axis of $22.00 \%$, and RMSE of $0.198$.
The $5$x$7$ beam splitter sample has a total efficiency of  $74.11 \%$, UE of $17.58 \%$, UE in off-axis of $14.07 \%$, and RMSE of $0.172$.
The properties based on the measurement are considerably close to the results obtained by calculation in the previous section.
Little discrepancies between the experimental and theoretical efficiencies are due in part to minor geometric imperfections in the fabricated device.
In general the diffraction efficiency in orders often strongly depends on the errors in fabrication processes, e.g., etching depth, feature width, slope steepness ,and feature rounding.
The comparison with theoretical and experimental efficiency of $7$x$7$ beam splitter and $5$x$7$ beam splitter are summarized in Fig. \ref{fig:experimental}(a) and (c), respectively.
To exclude the effects which may occur during measurement such as Fresnel reflection loss and power detector offset, the measured efficiency is normalized.
The results of the comparison show that the experimental results have a strong correlation with the designs.
Overall, the optimized samples display experimental performances which are significantly higher than the theoretical performances of initial devices before optimization.

\section{CONCLUSIONS}
\label{sec:conclusion}
We propose a general method for optimizing diffractive optical elements using the adjoint variable method.
Due to the adjoint method, we can compute the gradient of the figure of merit with respect to the design parameters efficiently even using rigorous diffraction theory, independently of the number of design parameters.
We show that our method can be applied to the inverse design of two-dimensional beam splitters: a wide-angle $7$x$7$ beam splitter with a $53^\circ$ diffraction angle as well as a beam splitter that generates a $5$x$7$ non-square array.
The optimized beam splitters show considerable improvement of uniformity while maintaining the initial diffraction efficiency. 
The experimental results obtained by the illumination of the fabricated optical elements using an incident laser of $940 nm$ wavelength with TE-polarization have been compared with the numerical results. 
As numerical simulation and experimental results were found to be in good agreement, our optimization method can be considered proven to be an effective design tool for wide-angle diffractive fan-out beam splitters.

\acknowledgments 
This research has received funding from the European Union’s Horizon 2020 research and innovation program under the Marie Sk\l odowska-Curie Grant Agreement No. 675745(MSCA-ITN-EID NOLOSS). 

\pagebreak

\bibliography{report} 
\bibliographystyle{spiebib} 
\vfill
\noindent \copyright (2020) COPYRIGHT Society of Photo-Optical Instrumentation Engineers (SPIE).
\end{document}